\documentclass[
    ,final            
  ]
  {aipproc}

\layoutstyle{8x11single}

\begin{document}

\title{Recent Results from HERA Experiments}


\author{Katarzyna Wichmann, on behalf of the H1 and ZEUS Collaborations}{
  address={Hamburg University, Luruper Chaussee 149, 22761 Hamburg, Germany}
}

\begin{abstract}
 Recent results from HERA are presented. The main reviewed subjects are 
polarized DIS cross sections, parton density determination, diffractive PDFs, 
multi-jet production and searches for physics beyond the Standard Model.
 
\end{abstract}

\maketitle


\section{HERA Physics}

HERA is an electron\footnote{``Electron'' refers both to electrons and positrons unless 
specified}-proton ($ep$) collider operating in DESY in Hamburg, Germany.
At HERA, electrons of energy $E_e = 27.6$ GeV collide with protons of energy $E_p = 820$ or 
$920$ GeV, corresponding to a center-of-mass energy  of $301$ or $319$ GeV, respectively. 

In autumn 2003 the HERA accelerator started operation of the second phase of its $ep$ collider 
programme, so called HERA II. 
The $e^+p$ and $e^-p$ data collected by the H1 and ZEUS experiments since then were taken with 
a longitudinally polarized positron beam for the first time.
At HERA transverse polarization of the positron beam arises naturally through synchrotron 
radiation via the Sokolov-Ternov effect~\cite{sokolov}. In 2000 a pair of spin rotators was 
installed in the beamline on either side of the H1 detector, allowing transversely polarized 
positrons to be rotated into longitudinally polarized states and back again. 
The degree of polarization is constant around the HERA ring and is continuously measured using 
two independent polarimeters~\cite{pol}.

H1 and ZEUS collaborations are colliding experiments located at HERA. 
Both have big, almost full-coverage detectors located at HERA interactions points.
A detailed description of the detectors can be found elsewhere~\cite{det}.

HERA is a multipurpose machine but it is mainly ideally suited for detailed studies of 
perturbative Quantum Chromodynamics (QCD) and for testing new QCD predictions. 
The HERA experiments extended the kinematic range by more than two orders of magnitude in both 
$x$ and $Q^2$ with respect to that accessible to the earlier, fixed target experiments and have a
common phase-space with the future LHC accelerator. 

This review presents only a subjective selection of the very broad spectrum of results
delivered by the H1 and ZEUS experiments at the HERA collider.

\section{Polarized Deep Inelastic Scattering Cross Sections}

For the neutral current (NC) and charged current (CC) Deep Inelastic Scattering (DIS) reaction, 
$ep \rightarrow  e^{'} + X$ and $ep \rightarrow  \nu + X$ respectively, 
the four-momentum transfer can be calculated as 
$Q^2 = -q^2 = -(k - k^{'})^2$, where $k$ and $k^{'}$ are the incident and the scattered lepton 
four-momentum, respectively. The fraction $x$ of the proton momentum carried by the struck 
quark is $x = Q^2/(2P \cdot q)$ with $P$ denoting the proton four-momentum. The inelasticity is
$y = (q \cdot P )/(k \cdot P )$ and $W = \sqrt{(q + P )^2}$ measures the energy of the 
hadronic system. 

The Standard Model (SM) predicts that the cross sections for 
charged and neutral current $ep$ DIS should exhibit specific dependencies on the longitudinal 
polarization of the incoming lepton beam. The absence of right-handed charged currents leads 
to the prediction 
that the CC cross section will be a linear function of polarization, 
vanishing for right-handed (left-handed) electron (positron) beams.

The H1 and ZEUS collaborations have measured the NC and CC DIS cross sections (total, 
differential and double differential) for electron and 
positron beams with positive and negative polarizations and compared results with the SM 
predictions~\cite{sigma_pol_h1, sigma_pol_zeus}.
At large $Q^2$ the structure function $F_3$, which provides information on valence quark
distributions, can be measured by subtracting the NC cross sections of
electron-proton and positron-proton DIS. Both Collaboration have extended the previous 
$F_3$ measurements~\cite{f3_h1,f3_zeus} using the HERA II data. A reduction of statistical 
error of these measurements is due mainly to much increased electron sample.
The measurements of the H1~\cite{sigma_pol_h1} and ZEUS~\cite{sigma_pol_zeus}
Collaborations are in good agreement. The expectation of the SM, evaluated with the 
H1 PDF 2000 and CTEQ6D PDF for H1 and ZEUS measurement, respectively, gives a good 
description of the data. 

\begin{figure}
  \includegraphics[height=.275\textheight]{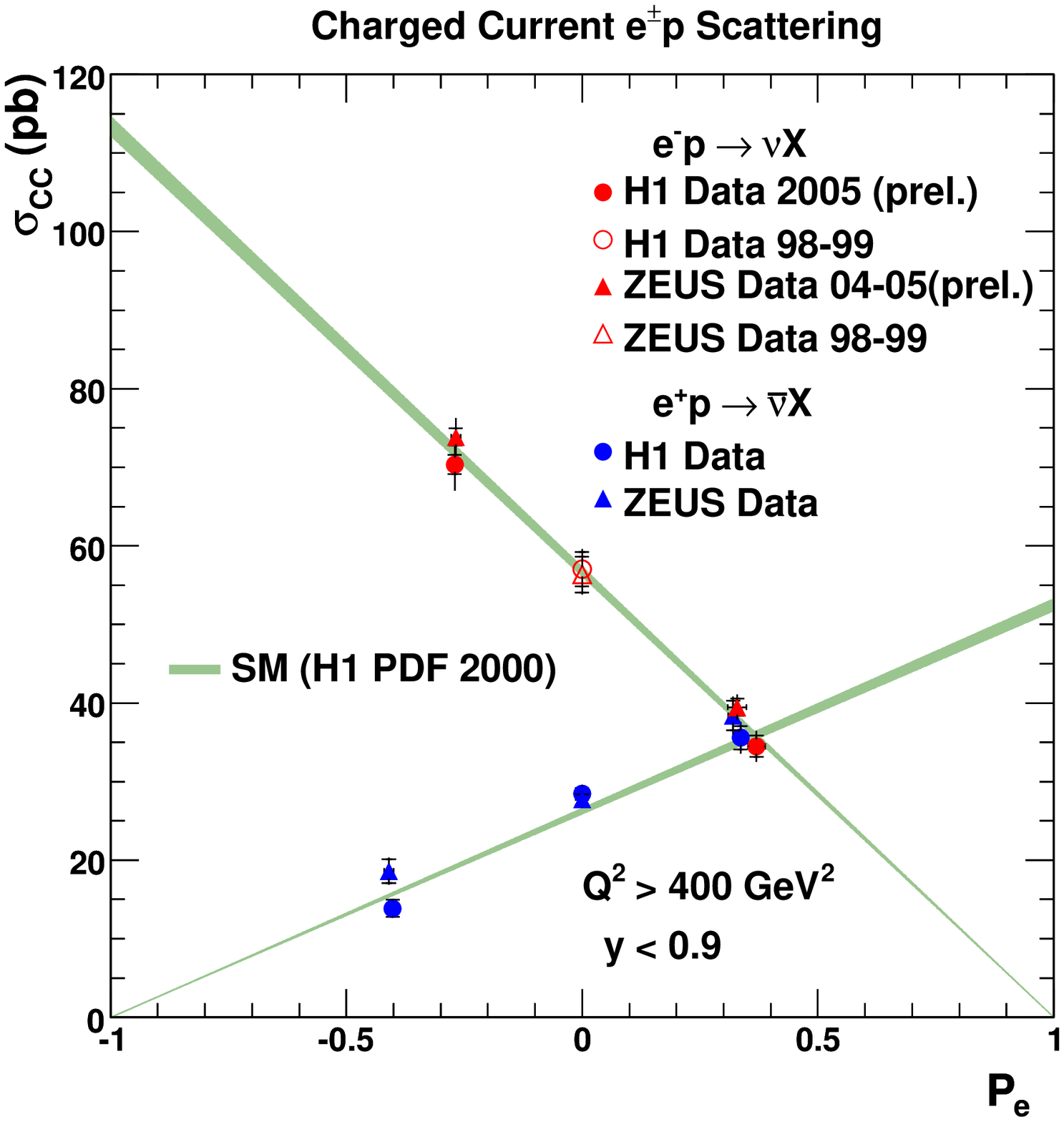}
 \includegraphics[height=.275\textheight]{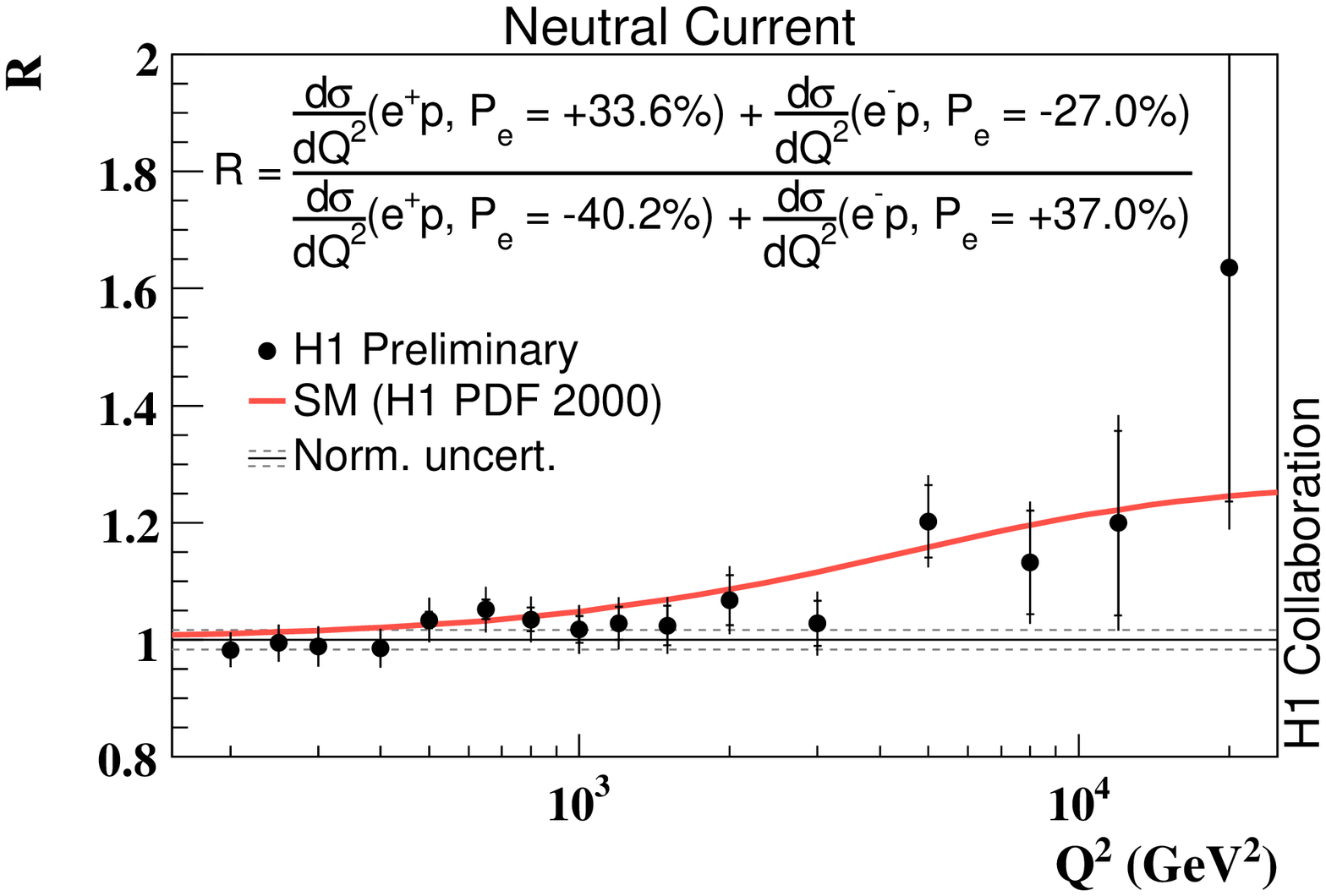}
  \label{cc-nc}
  \caption{Left: The total cross section for $e^-p$ and $e^+p$ CC DIS as a function of
the longitudinal polarization of the lepton beam measured by the H1 and ZEUS Collaborations.
The inner error bars represent the statistical uncertainties and the outer error bars 
represent the total error. The normalization uncertainty is not included on the error bars.
The lines show the predictions of the SM evaluated using the H1 PDF 2000.
Right: The $Q^2$ dependence of the combined $ep$ left- and right-handed NC cross section
ratio R. The data (solid points) are compared to the SM prediction.}
\end{figure}

Fig.~\ref{cc-nc} on the left shows CC DIS cross sections for electron and 
positron beams as measured by H1 and ZEUS experiments as a function of polarization. 
There is a clear linear dependence of the CC DIS cross section 
as a function of polarization. The CC cross section is consistent with zero for right-handed 
(left-handed) electron (positron) beams. 
The results are compared to the SM predictions and they agree well. 

Fig.~\ref{cc-nc} on the right shows the $Q^2$ dependence of the combined $ep$ left- 
and right-handed NC cross section ratio R, measured by the H1 experiment, compared to the 
SM prediction. There is a clear parity violation observed at high $Q^2$ which is in 
a good agreement with the SM prediction.

\section{QCD Fits with HERA Data}

The gluon density in the proton contributes only indirectly to the inclusive DIS cross sections.
However it makes a direct contribution to jet cross sections through
boson-gluon and quark-gluon scattering.
Tevatron high-$E_T$ data~\cite{highet_d0,highet_cdf} have been used to constrain
the gluon in the fits of MRST~\cite{mrst1,mrst2} and CTEQ~\cite{cteq}.
However, these data suffer from very large correlated
systematic uncertainties from a variety of sources.

In the ZEUS-JETS fit~\cite{zeus-jets},
ZEUS NC $e^+p$ DIS inclusive  jet cross sections~\cite{dis_jets}  and direct
photoproduction dijet cross sections~\cite{php_jets} have been used to constrain the gluon. 
These jet data together with ZEUS data on NC and CC $e^+p$ and $e^-p$
DIS inclusive cross sections were used as inputs to an next-to-leading order 
(NLO) QCD DGLAP analysis in order to determine the PDFs within a single experiment.
The predictions for the jet cross sections are
calculated to NLO in QCD and are used in the fit
rigorously, rather than approximately as in previous fits~\cite{mrst1,mrst2,cteq}.
The agreement between the MRST and CTEQ PDFs to the ZEUS-JETS PDFs, considering the size of the
uncertainties on each PDF set, is good.
The shapes
of the PDFs are not changed significantly by the addition of jet data. The decrease in
the gluon distribution uncertainty is significant. In the mid-$x$ range,
over the full $Q^2$ range a decrease in uncertainty by a factor of about two is found.

\begin{figure}
  \includegraphics[height=.275\textheight]{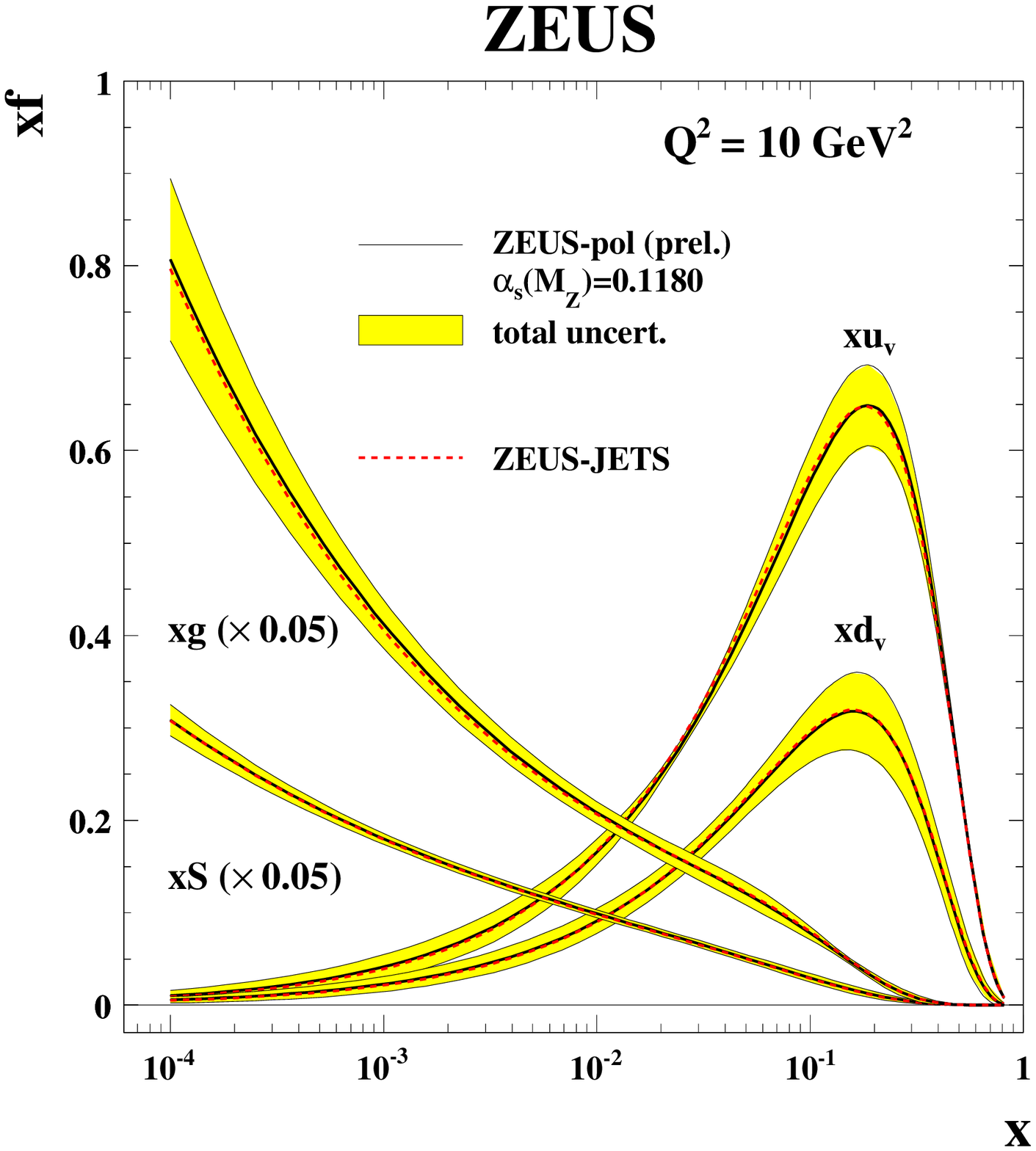}
  \includegraphics[height=.275\textheight]{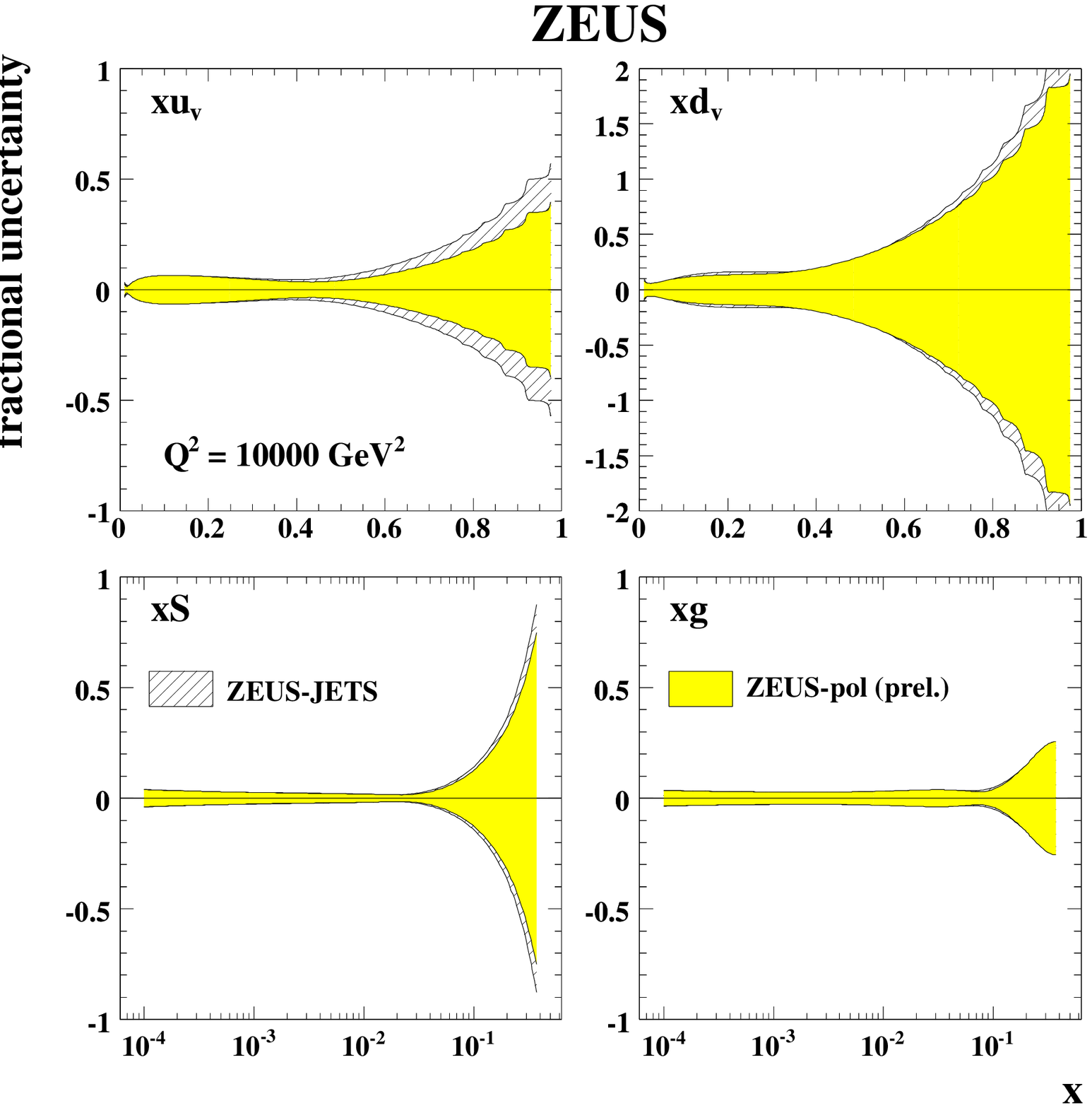}
  \label{qcd}
  \caption{Left: Comparison of the PDFs extracted from the ZEUS-pol PDF fit with those 
extracted in the ZEUS-JETS PDF analysis. The ZEUS-pol PDFs are shown with their total 
experimental uncertainty and the ZEUS-JETS PDFs are shown by their central values.
Right: Total experimental uncertainty on the PDFs at $Q^2 = 10,000$ GeV$^2$ for the 
ZEUS-pol fit (central error bands) compared to the total experimental uncertainty on 
the PDFs for the previous ZEUS-JETS fit (outer error bands). The uncertainties are shown
as fractional differences from the central values of the fits. The total experimental 
uncertainty includes the statistical, uncorrelated and correlated systematic uncertainties
and normalizations for both fits.}
\end{figure}

A new PDF fit has been done including new $e^-p$ NC and CC inclusive 
cross-sections~\cite{sigma_pol_zeus} taken with polarised electron beams~\cite{qcd-ew}. 
These data are all at high $Q^2 > 200$ GeV$^2$ and thus they give an improved 
determination of the PDFs at high-$x$, particularly for the $u$-valence distributions, 
as the cross sections for NC and CC  $e^-p$ scattering are both sensitive to the 
$u$-valence PDF at high-$x$.
The PDFs for the ZEUS-pol fit are shown in Fig.~\ref{qcd} left, compared to those
of the ZEUS-JETS fit, at $Q^2 = 10$ GeV$^2$. The shapes of the PDFs are not changed 
significantly by the addition of polarized data. In Fig.~\ref{qcd} right, the 
uncertainty of the PDFs for fits with and without the new polarized data are shown. 
The precision of the high-$x$ PDF distributions is improved in comparison to the ZEUS-JETS
fit, in particular for the $u$-valence distributions. The precision of the $u$-valence PDFs 
is now compatible to that of global fits which include fixed target data as well~\cite{fixed}.
This figure shows the PDF uncertainties at a scale $Q^2 = 10,000$ GeV$^2$, where the PDF
uncertainties are a vital input to the estimate of precision $W^{\pm}$ and $Z^0$ 
cross-sections at the LHC.

\begin{figure}
  \includegraphics[height=.275\textheight]{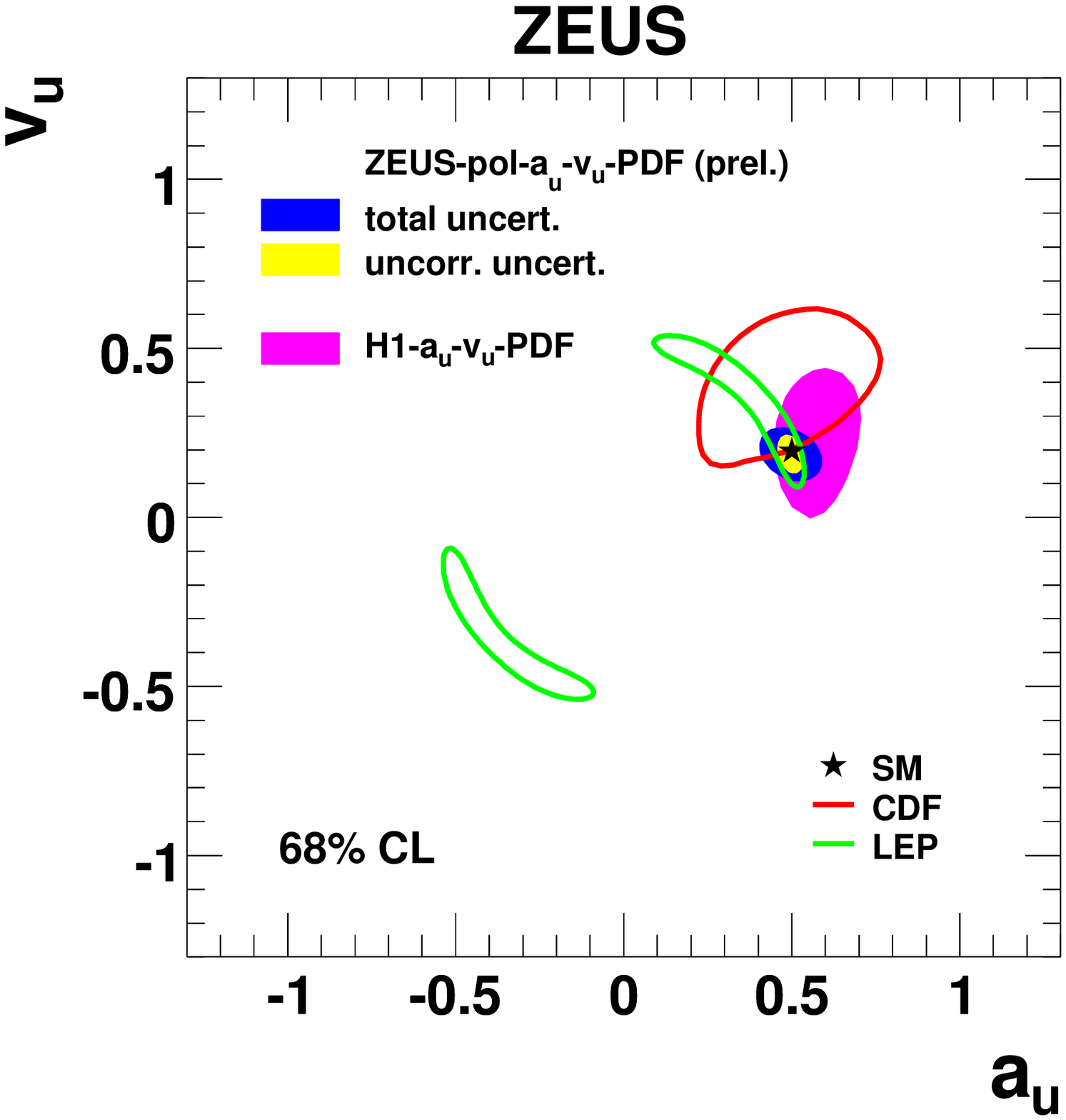}
 \includegraphics[height=.275\textheight]{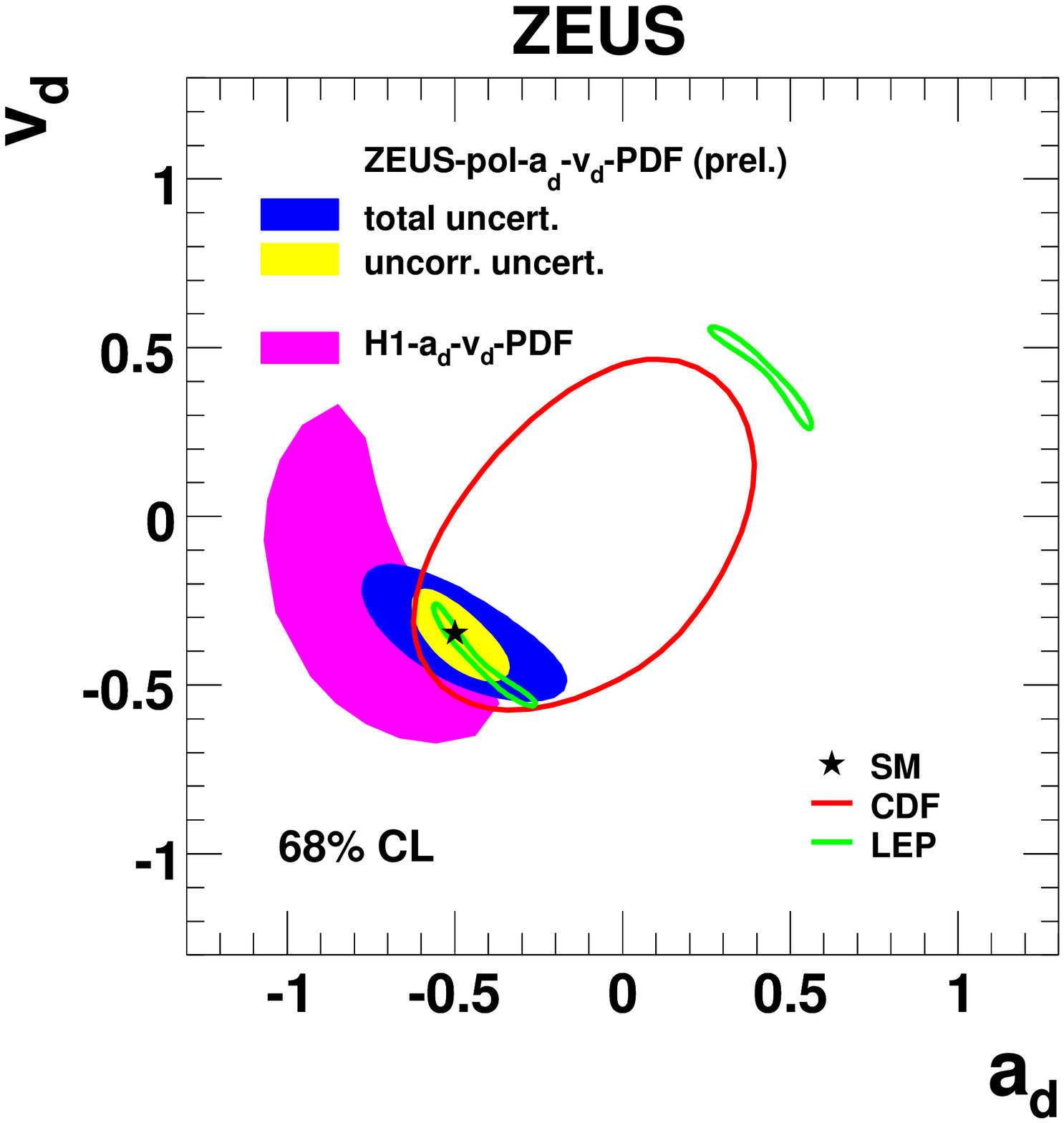}
  \label{ew}
  \caption{The 68\% confidence level contours for the electroweak parameters $a_u$ vs $v_u$
(left) and $a_d$ vs $v_d$ (right) from the ZEUS-pol-$a_u-v_u$ and the 
ZEUS-pol-$a_d-v_d$ fit, compared to the same contours extracted by other experiments.}
\end{figure}

The present HERA data give sufficient precision to make an attempt to constrain the 
electro-weak (EW) sector of the SM. The CC cross section depends strongly on 
the $W$ propagator and this dependence is used to extract $M_W$ and $G_F$, 
in a space-like process at high $Q^2 \approx 400$ GeV$^2$. The weak vector and axial-vector 
neutral current couplings of quarks, $v_u, a_u, v_d, a_d$, can also be determined.
The ZEUS collaboration performed a combined QCD and EW analysis~\cite{qcd-ew} 
of the data.
The H1 collaboration has made a similar analysis~\cite{qcd-h1} of HERA I data. The new ZEUS
analysis extends the results by using the information from polarized lepton beams in the
HERA II data. The left plot of Fig.~\ref{ew} compares the contour of $a_u$ vs $v_u$ from
this analysis to the H1 analysis. The contours obtained by CDF~\cite{ew_cdf} and by 
LEP~\cite{ew_lep} in the light quark sector are also shown for comparison. In the right 
plot the same comparison is shown for the contour $a_d$ vs $v_d$. It can be seen very 
clearly that the new polarized data make a significant impact on the estimation of the 
weak NC vector couplings in the light quark sector. 

\section{Diffractive PDFs from HERA}

\begin{figure}
\includegraphics[height=.275\textheight]{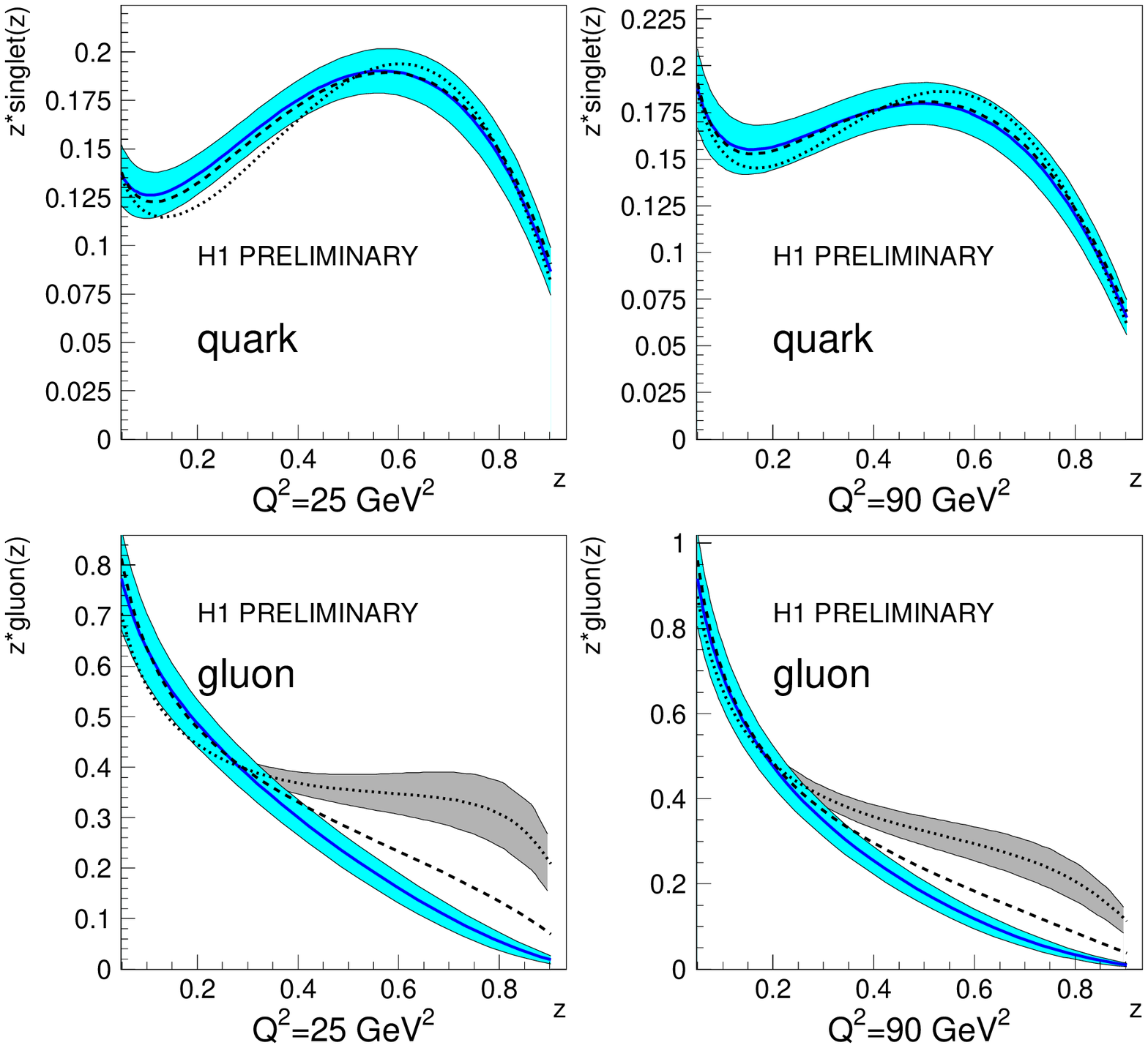}
\includegraphics[height=.275\textheight]{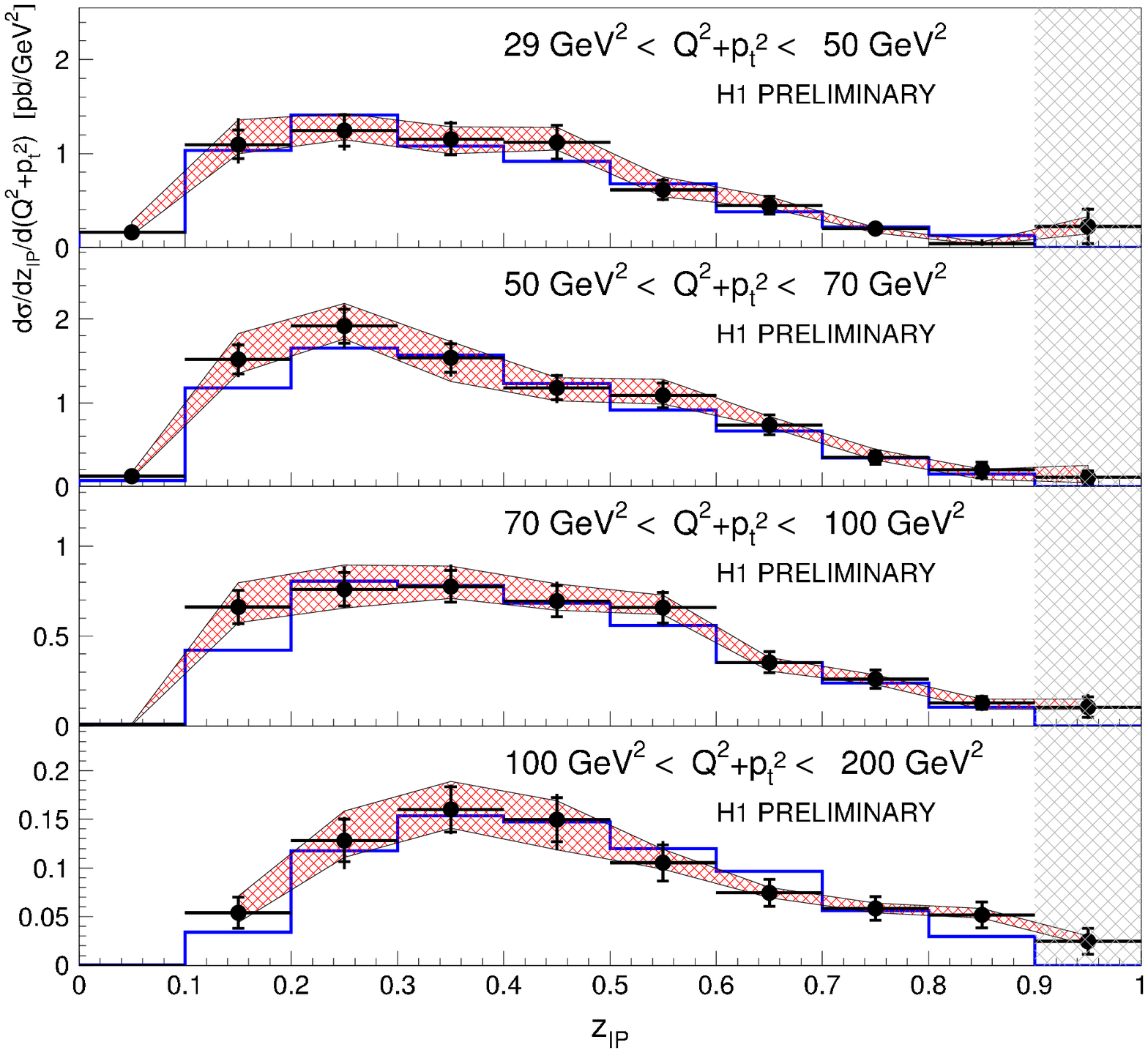}
  \label{diff}
  \caption{Left: The diffractive singlet density (top) and diffractive gluon density 
(bottom) for two values of the hard scale $\mu=25$ GeV$^2$ (left) and $\mu=90$ GeV$^2$ (right). 
The blue line indicates the combined fit, surrounded by the experimental uncertainty 
band in light blue. The two dashed lines show the predictions of the H1 2006 DPDF 
fit~\cite{h1-dpdf} for comparison.
Right: Cross section of diffractive dijets doubly differential in $z_{IP}$ and the scale 
$\mu = Q^2 + p_{t}^2$. The data are shown as black points with the inner and outer error bar 
denoting  the statistical and uncorrelated systematic uncertainties respectively. 
The red hatched band indicates the correlated systematic uncertainty. 
The blue line shows the NLO QCD prediction based on the combined fit. 
Data points in the grey hatched area were not included in the fit due to problems with 
the hadronization corrections.}
\end{figure}

The study of hard diffraction processes at HERA offers an important insight on the 
diffraction and the 'Pomeron' structures. 
Hard diffractive processes (inclusive deep inelastic scattering (DIS),
jet production and production of charmed quarks) can be well described by factorizing 
the cross section into a Pomeron flux and a hard QCD scattering process with a parton in 
the 'Pomeron' which is described by diffractive parton densities (DPDF). 
Diffractive parton densities have been determined in DGLAP QCD analysis using inclusive 
diffractive HERA data~\cite{diff1,diff2} and have been found to be dominated by the 
gluon distribution. 
Diffractive dijet production is directly sensitive to the gluon component of the 
diffractive exchange and for DIS~\cite{diff3} is in reasonable agreement with the QCD fits 
to the inclusive diffractive data. 
The H1 Collaboration has performed a new measurement of diffractive dijet cross sections in 
DIS~\cite{diff4} with the integrated luminosity increased by a factor 5 with respect to previous 
results~\cite{diff3}. 
The diffractive quark and gluon distributions have been determined from a combined NLO QCD 
fit performed by the H1 Collaboration to the differential dijet cross sections and the 
inclusive diffractive structure function $F_2^{D(3)}$~\cite{diff4}.
The diffractive gluon and quark singlet distributions are shown in the left plot of 
Fig.~\ref{diff} for a hard scale $\mu^2 = 25$ GeV$^2$ and $\mu^2 = 90$ GeV$^2$. 
The error bands indicate the preliminary systematic experimental errors. 
The combined fit for the first time constrains both the diffractive gluon and quark 
densities remarkably well in the range $0.05 < z_{IP} < 0.9$, where $z_{IP}$ is a longitudinal
momentum fraction of a parton entering hard sub-process w.r.t diffractive exchange. 
An example of dijet cross sections compared to the predictions based on the combined 
fit are shown in the right plot of Fig.~\ref{diff}. 
The overall agreement for all distributions is reasonable, showing that a consistent 
NLO QCD prediction is possible by a suitable choice of 
the diffractive parton distributions.

\section{Multi-jet Production at HERA}

\begin{figure}
\includegraphics[height=.265\textheight]{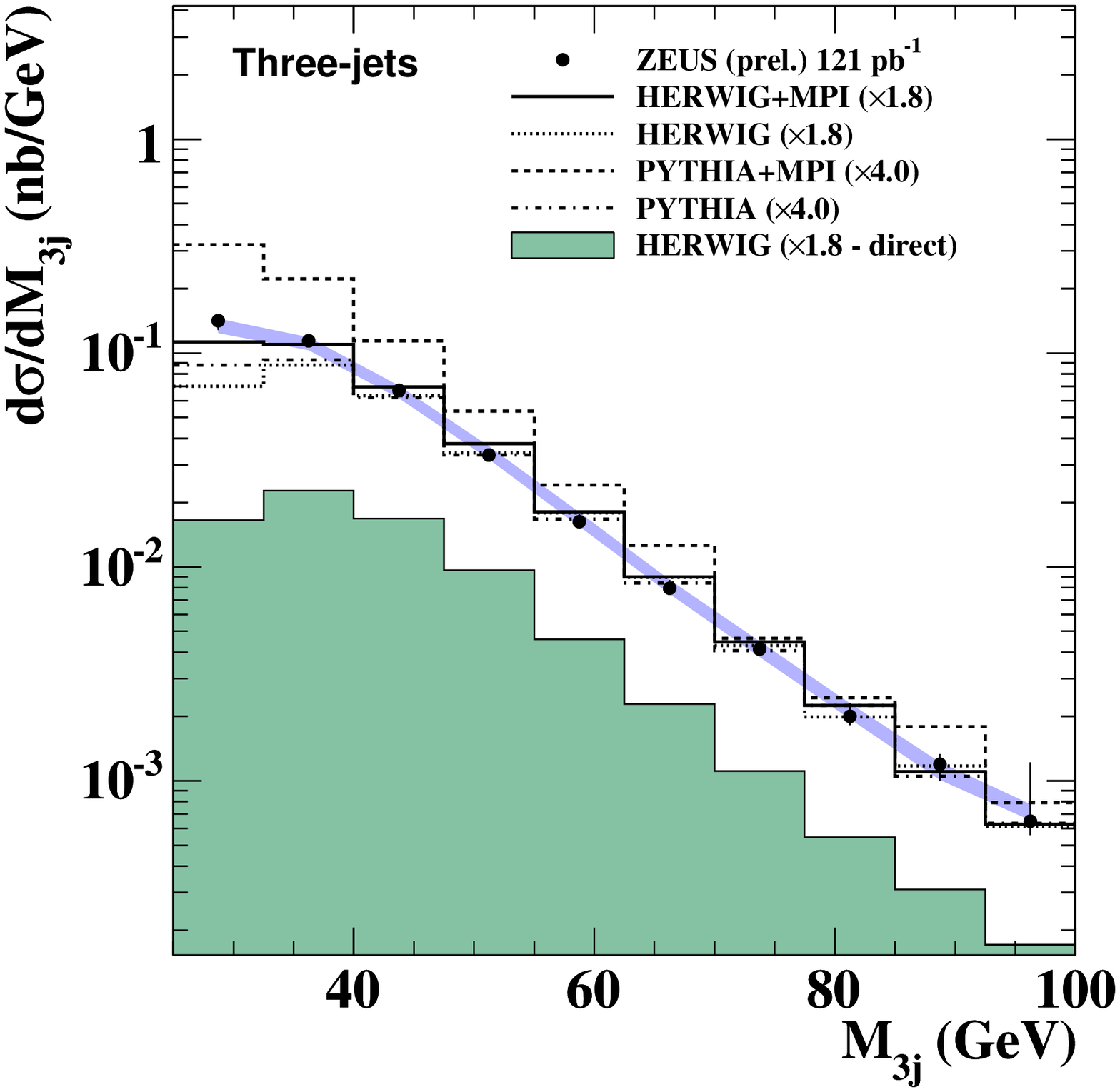}
\includegraphics[height=.265\textheight]{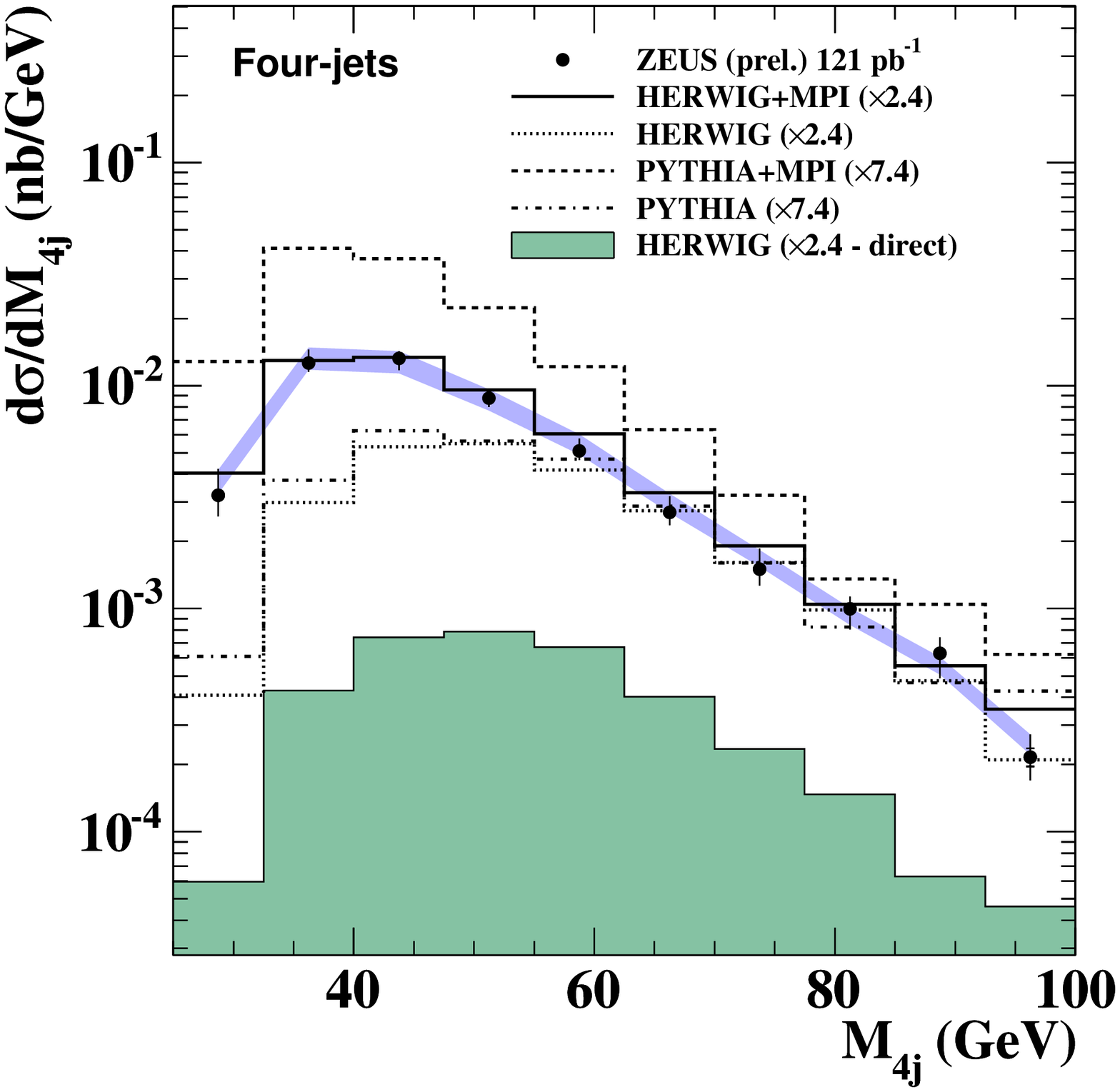}
  \label{jets}
  \caption{The multi-jet cross section in the semi-inclusive mass for three- (left) and 
four-jet sample (right). The calorimeter energy scale uncertainty is shown as the shaded band.
Also shown are the {\tt HERWIG} and {\tt PYTHIA} predictions, with and without MPIs, 
as well as the {\tt HERWIG} direct component. Each MC cross section has been scaled by 
the amount indicated in the legend.}
\end{figure}

Three- and four-jet states have been measured by the ZEUS Collaboration in 
photoproduction at HERA~\cite{mpi}. 
The three-jet measurement has extended the phase-space of previous results 
and is based on over seven times more luminosity compared to the previous HERA measurements.
The four-jet photoproduction cross section has been measured for the first time and 
represents the highest order processes studied at HERA.
Both the three- and four-jet events 
have been studied in a semi-inclusive, $M_{nj}>  25$ GeV, and high-mass region, 
$M_{nj}< 50$ GeV. 
In photoproduction multi-jet events are of particular interest since they are mainly
produced by processes of high order in the strong coupling constant, $\alpha_s$. 
Perturbative quantum chromodynamics calculations for multi-jets in photoproduction are 
only available at $O(\alpha\alpha_s^2)$ for the hard matrix elements. 
At HERA in resolved photoproduction multi-parton interactions (MPIs) can occur.
In MPI models, more than one pair of partons from the incoming hadrons may interact.
The secondary scatters generate additional hadronic energy flow in the event, 
the topology of which is poorly understood theoretically. 
The three-jet and four-jet cross sections are reasonably described by 
both {\tt PYTHIA} and {\tt HERWIG}, run without MPIs, in the high-mass region. 
In the semi-inclusive-mass region, the MC without MPIs underestimated the data. 
When MPIs were added to the MC simulations, the agreement between the models and data 
was generally improved, although the Pythia model, tuned to generic collider data, 
augmented the cross section too much. As an example Fig.~\ref{jets} shows the three- 
and four-jet cross section measure differentially in the multi-jet mass, $M_{nj}$,
compared to  the {\tt HERWIG} and {\tt PYTHIA} models with and without MPIs. Both MC 
models describe the ${\rm d}\sigma/{\rm d}M_{jn}$ cross sections well at high $M_{nj}$
but significantly underestimate them at lower values. 

\section{High-$P_T$ Isolated Leptons at HERA}

At HERA isolated leptons arise in events with a topology matching the electron or muon decay 
channel of singly produced W bosons. 
Single W production is a rare SM process and an important source of background 
to searches for physics beyond the Standard Model at HERA~\cite{iso1,iso2}. 
Investigations of the process $ep \rightarrow eW X, W \rightarrow l\nu$, 
where $l = \mu, e, \tau$, have been performed at HERA by both the H1~\cite{iso1,iso3,iso4} 
and ZEUS~\cite{iso5} collaborations. 
The H1 collaboration observes an excess of events with isolated muons or electrons, 
high missing transverse momentum and large values of hadronic 
transverse momentum over the SM prediction, dominated by single W production. 
The ZEUS results based on searches for isolated electrons and muons at a center-of-mass 
energy of 300 GeV and 318 GeV do not confirm this excess. 
Both collaboration completed the investigation using HERA I and HERA II data up to the 
year 2005. No excess over the Standard Model predictions is observed by ZEUS. 
Tab.~\ref{iso} summarized the results for the H1 search, both for electron and positron data
for the $P_T^X > 25$ GeV, where $P_T^X$ is a transverse momentum of the hadronic system from
the single $W$ production at HERA.  
The H1 collaboration finds an excess of data events over the SM prediction for the 1994-2004 
$e^+p$ data sample. The excess is not present in the $e^-p$ sample. Due to different 
phase-space and $W$ efficiencies in the H1 and ZEUS measurements the results can not be directly
compared. There is a combined effort of both experiment to have comparable results - the 
measurement of isolated leptons with high transverse missing momentum might be the only
chance for a discovery at HERA.

\begin{table}
\begin{tabular}{cccc}
\hline
\tablehead{1}{c}{b}{$P_T^X > 25$ GeV}&
\tablehead{1}{c}{b}{$e$ channel (data/SM)}&
\tablehead{1}{c}{b}{$\mu$ channel (data/SM)}&
\tablehead{1}{c}{b}{Combined $e$ and $\mu$ channels (data/SM)}   \\
\hline
Electrons, 98-05, 121 pb$^{-1}$  & $2/2.4 \pm 0.5$ & $0/2.0 \pm 0.3 $ & $2/4.4 \pm 0.7$\ \ \ \ \\
Positrons, 94-04, 158 pb$^{-1}$  & $9/2.3 \pm 0.4$ & $6/2.3 \pm 0.4$  & $15/4.6 \pm 0.8$ \\
\hline
\end{tabular}
\caption{Results from H1 Collaboration search for events with isolated leptons and large
missing transverse momentum in the final state and with large transverse momentum of the 
hadronic system $P_T^X > 25$ GeV. The numbers for the data are compared to 
the SM predictions.}
\label{iso}
\end{table}

\bibliographystyle{aipproc}   

\end{document}